\documentclass{elsart3p}
\usepackage{graphicx,epsfig,natbib}
\usepackage{amsmath}

\newcommand{\apj}{ApJ}
\newcommand{\aap}{A\&A}
\newcommand{\apjl}{ApJL}
\newcommand{\mnras}{MNRAS}
\newcommand{\aj}{AJ}
\newcommand{\apjs}{ApJS}

\newcommand{\physrep}{Physical Review}
\newcommand{\hs}{\hspace{1mm}}
\def\lsim{\mathrel{\raise.3ex\hbox{$<$\kern-.75em\lower1ex\hbox{$\sim$}}}}
\def\gsim{\mathrel{\raise.3ex\hbox{$>$\kern-.75em\lower1ex\hbox{$\sim$}}}}
\begin{document}
\begin{frontmatter}
\title{Requirements for Cosmological 21cm Masers}
\journal{New Astronomy}

\author[Dijkstra]{Mark Dijkstra\corauthref{cor}\thanksref{now}},
\corauth[cor]{Corresponding Author}
\thanks[now]{Present address: Astronomy Department, Harvard University, 60 Garden Street, Cambridge, MA 02138, USA}
\ead{mdijkstr@cfa.harvard.edu}
\address[Dijkstra]{School of Physics, University of Melbourne, Victoria 3010, Australia}

\author[Loeb]{Abraham Loeb}
\ead{aloeb@cfa.harvard.edu}
\address[Loeb]{Astronomy Department, Harvard University, 60 Garden Street, Cambridge, MA 02138, USA}
\begin{abstract}
We perform Monte-Carlo calculations of the radiative transfer of Ly$\alpha$ photons emitted by a source embedded in a neutral collapsing gas cloud. This represents a young galaxy or quasar during the early stages of the epoch of reionisation (EoR). After computing the Ly$\alpha$ spectrum as function of radius and time, we find that the Ly$\alpha$ color temperature may be negative in large volumes surrounding the central source. Motivated by this result, we explore the prospects for a population inversion in the hyperfine levels of atomic hydrogen via the Wouthuysen-Field (WF) effect. The reason for this exploration is clear: if 21-cm masers exist during the EoR, they could greatly boost the expected 21-cm flux from this epoch. We find that population inversion is unlikely to occur for several reasons: (1) the required Ly$\alpha$ luminosities of the central source exceed $\sim 10^{45}$ erg s$^{-1}$. The radiation pressure exerted by such a large Ly$\alpha$ flux likely halts the collapse of the cloud; (2) When quantum corrections to the WF-coupling strength are applied, the required Ly$\alpha$ luminosities are (even) larger by orders of magnitude; (3) A relatively low flux of Ly$\alpha$ photons that is produced via other channels (x-ray heating, collisional excitation of hydrogen) prevents the Ly$\alpha$ color temperature from becoming negative.
\end{abstract}
\end{frontmatter}
\section{Introduction}
\label{sec:intro}

Detecting the redshifted HI 21-cm line from the epoch of reionisation (EoR) is
among the greatest observational challenges in present-day
cosmology. Experiments such as the Low Frequency Array
(LOFAR)\footnote{www.lofar.org}, the Primeval Structure Telescope 
\citep[PAST,][]{Pe06} and the Murchinson Widefield Array
(MWA)\footnote{http://www.haystack.mit.edu/ast/arrays/mwa/Site/index.html},
are currently being developed with the goal of detecting this EoR 21-cm signal within the next few years.

The strength and sign of the 21-cm signal depend on the spin temperature of
the hydrogen gas, $T_s$, which is defined through the ratio of atoms in the
ground and excited states of the hyperfine transition in the electronic
ground state of hydrogen, $n_2/n_1 \equiv g_2/g_1$ exp$[-T_*/T_s]$, where
the ground and excited states are denoted by the labels '1' and '2',
respectively and $T_*=0.068$ K. If we denote the temperature of the Cosmic
Microwave Background (CMB) by $T_{\rm CMB}$, then neutral hydrogen gas can
be seen in emission if $T_s > T_{\rm CMB}$. When $T_s < T_{\rm CMB}$, the
neutral hydrogen gas can be seen in absorption against the CMB.

The distribution of atoms in the hyperfine levels is affected by collisions
between atoms, absorption and re-emission of the Cosmic Microwave Background (CMB) photons, and absorption and re-emission of Ly$\alpha$ photons \citep{review}. These Ly$\alpha$ photons were either emitted by nearby sources, produced as local recombination emission, or in cascades following photo-excitation by photons with energies $E\in [10.2-13.6]$ eV \citep{PF06a}. When the Ly$\alpha$ scattering rate dominates the collisional and CMB scattering rate, the hydrogen spin temperature, $T_s$, approaches the Ly$\alpha$ color temperature, $T_{\alpha}$ \citep{Wouthuysen52}. The color temperature has been shown to approach the gas temperature, $T_{\alpha}\rightarrow T_{\rm gas}$,  because of the frequency redistribution that Ly$\alpha$ photons experience while resonantly scattering through an optically thick medium \citep{Field59}. Because of this 'Wouthuysen-Field' effect, it is commonly assumed that the presence of Ly$\alpha$ photons strengthens the coupling of the spin to the gas temperature \citep[for the most detailed calculation see][]{Hirata06} and that neutral hydrogen gas is detectable in emission.\\

In this paper, we examine whether there are physical conditions under which the
WF-effect produces population inversion (i.e. $T_s <0$) in the hyperfine
levels. This investigation is motivated by recent Monte-Carlo calculations of Ly$\alpha$ radiative transfer, which show that Ly$\alpha$ photons emerge from neutral collapsing gas clouds with a systematic blueshift \citep[e.g.][]{Zheng02,Dijkstra06}. This blueshift was found to increase with Ly$\alpha$ optical depth and with the infall speed of the gas \citep{Dijkstra06}. In this paper we show that this may lead to a negative Ly$\alpha$ color temperature, and potentially to a negative spin temperature. The possibility of population inversion via the WF-effect and the accompanying stimulated 21-cm emission has been addressed in previous work \citep[e.g.][]{Va67,S67,V68,Kaplan79}, but (obviously) not in the light of the results of these recent radiative transfer calculations, and not in the context of the EoR \footnote{\citet{SN97} have investigated whether hydrogen masers can exist during EoR. However, they focused on hydrogen  $n\alpha$ recombination lines, which correspond to the transitions of the form $n\rightarrow n-1$, where $n$ is the principal quantum number of atomic hydrogen.  \citet{SN97} found that maser lines may be produced for $n\sim 120$ along the edges of early HII regions.}. Exploring in more detail whether $T_s<0$ barely needs justification: if high redshift 21-cm masers exist, they could boost the 21-cm EoR signal by orders of magnitude. Interestingly, \cite{CS06} recently showed that scattering of Ly$\alpha$ and Ly$\beta$ photons by deuterium can cause its spin temperature to be negative. Technically, this produces a 92-cm maser, although the associated gain is too small to provide a dramatic effect.

Before proceeding, we point out that \citet{Deguchi85} found that Ly$\alpha$ scattering in expanding gas clouds results in $T_s \rightarrow T_{\rm gas}$. One expects the same to apply to a collapsing gas cloud. However, the analysis presented in \citet{Deguchi85} focused on Ly$\alpha$ that was generated throughout the cloud. On the other hand, the young pre-reionisation universe may contain bright Ly$\alpha$ sources, such as young galaxies and/or (mini)-quasars, that are surrounded by cold, neutral, collapsing gas. In such scenarios the Ly$\alpha$ emission is concentrated strongly towards the center of the collapsing cloud. This introduces an important difference with the calculations of \citet{Deguchi85}. Photons that are emitted in the center of the cloud must propagate through a larger column of collapsing material. This causes more energy to be transferred from the infalling gas to the photons, which results in a larger overall blue-shift of the Ly$\alpha$ line \citep{Dijkstra06}, and we will show in this paper that this large blueshift is the main reason that the WF-effect may cause a population inversion. 

The outline of this paper is as follows. In \S~\ref{sec:gain} we discuss the expected maser amplification associated with a given negative hydrogen spin temperature.
In \S~\ref{sec:color} we derive an expression for the Ly$\alpha$ color temperature, which can be applied to arbitrary Ly$\alpha$ frequency distributions, and its relation to the hydrogen spin temperature.
In \S~\ref{sec:collapse} we discuss an example of a Ly$\alpha$ source surrounded by cold collapsing gas, in which the Ly$\alpha$ color temperature is negative.
This is followed in \S~\ref{sec:pressure}-\S~\ref{sec:sourcereq} by a
discussion of the factors that determine whether 21-cm masers can actually exist inside such clouds, before presenting our conclusions in \S~\ref{sec:conclusion}.
The parameters for the background cosmology used throughout this paper
are $\Omega_m=0.3$, $\Omega_{\Lambda}=0.7$, $\Omega_b=0.044$, and $h=0.7$
\citep{Spergel03}.

\section{The Gain Factor of a 21-cm Maser}
\label{sec:gain}
Stimulated emission of 21-cm emission occurs when $T_s <0$. Here, we
estimate the expected gain in the total 21-cm
emissivity. The change of the intensity in the line center of the 21-cm
line, $I$, with distance $s$, is given by \citep{RL79}
\begin{equation}
I(s)=I(0)\hs{\rm e}^{-\alpha s}+\frac{j}{\alpha}(1-\hs{\rm e}^{-\alpha s}),
\label{eq:radtrans1}
\end{equation} where $\alpha$ is the opacity of the gas at $\lambda_S=$21cm and $j$ is its volume emissivity. Furthermore, $I(0)$ is the intensity of the 21-cm photons before they enter the
gas, $I(0)=2k_BT_{\rm CMB}/\lambda_S^2$.  The 21-cm volume emissivity, $j$, and opacity, $\alpha$, are given by
\begin{equation}
 j=\frac{h\nu_S}{4\pi \Delta \nu_S}n_2 A_{21}
\label{eq:alpha}
\end{equation}
\begin{equation}
\alpha=\frac{h\nu_S\hs B_{21}}{4 \pi \Delta \nu_S}(3n_1-n_2)\\ \nonumber, 
\end{equation} where $\nu_S=1.4$ Ghz, is the frequency of the hyperfine transition; $\Delta \nu_S=\nu_S v_{\rm th}/c$, in which $v_{\rm th}$ is the thermal
velocity dispersion of the atoms; $A_{21}$ and $B_{21}$ are the Einstein
coefficients associated with the hyperfine transition. Eq.~(\ref{eq:radtrans1}) can be rewritten as 
\begin{equation}
I(s)=I(0)e^{s/R_M}-\frac{2k_BT_s}{\lambda_S^2}\big{(}e^{s/R_M}-1\big{)}
\label{eq:radtrans2},
\end{equation} where $R_M \equiv-1/\alpha$. We used $B_{21}=A_{21} c^2/2h\nu_0^3$, $n_1\sim n_H/4$ and the definition of the spin temperature (\S~\ref{sec:intro}) to rewrite $j/\alpha=2k_BT_s/\lambda_S^2$. Furthermore, we used that $T_*\ll |T_{\rm s}|$, which is valid throughout this paper. Since for a negative spin temperature $\alpha<0$, $I(s)$ increases exponentially with $s$. $R_M$ is the length scale over which $I(s)$ increases by one factor of $e$ and can be written as
\begin{eqnarray}
\label{eq:gain}
R_M\equiv- \frac{1}{\alpha}&\approx& 0.7
\Big{(}\frac{1+z}{11}\Big{)}^{-3}
\Big{(}\frac{1+\delta}{20}\Big{)}^{-1} \nonumber \\  &\times&
\Big{(}\frac{T_s}{-10 \hs K}\Big{)} \Big{(}\frac{T_{\rm gas}}{10 \hs
K}\Big{)}^{1/2}\hs {\rm kpc}.
\end{eqnarray}
Thus a reasonably small region in which $T_s<0$, also known as the ``masing region'', could produce a significant boost in $I(0)$. According to Eq.~(\ref{eq:gain}), the total maser amplification
increases toward lower $T_{\rm gas}$ and $\vert T_s\vert$ values and higher
densities.

\section{The Ly$\alpha$ Color and Hydrogen Spin  Temperature}
\label{sec:color}
\subsection{The Ly$\alpha$ Color Temperature}
\label{sec:lyacolor}
\citet{Madau97} defined the color temperature as
\begin{equation}
\frac{k T_{\alpha}}{h}=-\Big{[}\frac{\partial \log {\mathcal N_{\nu}}}{\partial \nu}\Big{]}^{-1}\approx-\Big{[}\frac{1}{J_{\alpha}(\nu)}\frac{\partial  J_{\alpha} (\nu)}{\partial \nu}\Big{]}^{-1},
\label{eq:tammr}
\end{equation} 
where $\mathcal{N}_{\nu}=c^2 J_{\alpha}(\nu)/2h \nu^3$ is the photon
occupation number. The more general expression for the Ly$\alpha$ color temperature is (Appendix~\ref{app:6level})
\begin{equation}
\frac{k T_{\alpha}}{h}=-\frac{ \int J_{\alpha} (\nu)\phi(\nu)
d\nu}{\int\frac{\partial J_{\alpha} (\nu)}{\partial \nu}\phi(\nu) d\nu}
=\frac{ \int J_{\alpha} (\nu)\phi(\nu) d\nu}{\int \frac{\partial \phi
(\nu)}{\partial \nu} J_{\alpha}(\nu) d\nu}
\label{eq:tcolor}
\end{equation} 
(also see Meiksin, 2006, Eq.~16). Here, $\phi(\nu)$ is the normalised line profile, $\int \phi(\nu) d\nu=1$ \citep[e.g.][]{RL79}. Frequency is expressed in terms of a dimensionless frequency variable, $x \equiv (\nu-\nu_0)/\Delta \nu_D$, where $\Delta \nu_D=\nu_0 v_{\rm th}/c$, in which $\nu_0=2.46\times 10^{15}$ Hz is the Ly$\alpha$ frequency. A negative/positive value of $x$ corresponds to a Ly$\alpha$ photon that is redshifted/blueshifted with respect to the line center, respectively. In the `core' of the line, $\phi(x) \sim e^{-x^2}$, while in the `wing' $\phi(x) \sim a/[\sqrt{\pi}x^2]$. Here, $a$ is the Voigt parameter, which is given by $a=A_{\alpha}/4 \pi \Delta \nu_D$ $=4.7\times 10^{-4}$ $(13 \hs {\rm km \hs s}^{-1}/v_{th})$, where $A_{\alpha}$ is the Ly$\alpha$ Einstein-A coefficient. The transition between core and wing is defined to occur at the frequency at which $e^{-x_t^2}=a/[\sqrt{\pi}x_t^2]$. Since the function $\phi(x)$ is strongly peaked around the line center, photons in this frequency range usually dominate the contribution to the integral.

At first glance the color temperature becomes negative when there are more photons with $\nu> \nu_0$ than with $\nu < \nu_0$, because $\partial \phi/\partial \nu <0 $ for $\nu > \nu_0$. This may occur in gas clouds moving towards a Ly$\alpha$ source, as these would 'see' a blueshifted Ly$\alpha$ line \citep{S67}. As was mentioned in \S~\ref{sec:intro} however, repeated scattering of photons in the line core rearranges them in frequency such that $T_{\alpha}\rightarrow T_{\rm gas}$ \citep[][]{Field59,V68}. This rapid rearrangement of photons in frequency space does not occur for photons that are far in the wing of the line profile, for which $|x| \gg 1$. Here, the probability that a photon is scattered from frequency $x$ back into the line core in a single scattering event is extremely small \citep[e.g.][]{Hummer62}. Instead, scattering pushes the photons back to the line center by an average amount of only $-1/x$ \citep{Osterbrock62}. Therefore, once a photon finds itself far in the line wing, it can stay there for $\sim x^2$ scattering events \citep{Adams72}. It follows that for a population of Ly$\alpha$ photons that are all in the blue wing of the line that: (i) $T_{\alpha}<0$, and (ii) scattering does not immediately redistribute the photons in frequency space such that $T_{\alpha} \rightarrow T_{\rm gas}$. Therefore, {\it a severely blueshifted Ly$\alpha$ line has the potential to sustain a negative color temperature while it interacts with the gas.} This negative color temperature may lead to a negative spin temperature, as we discuss next.

\subsection{The Hydrogen Spin Temperature}

\label{sec:tspin}
The expression for the equilibrium Hydrogen spin temperature is
(see Appendix~\ref{app:6level})
\begin{equation}
T_s=\frac{T_{\rm cmb}+y_\alpha T_{\alpha}+y_cT_{\rm gas}}{1+y_{\alpha}+y_c},
\label{eq:tspin}
\end{equation} 
where
$y_{\alpha}=\frac{4P_{\alpha}T_{*}Q[J_{\alpha}(\nu)]}{27A_{21}T_{\alpha}}$
and $y_{c}=\frac{C_{21} T_*}{A_{21} T_{\rm gas}}$. Here, $A_{21}$ and
$B_{21}$ are the Einstein rate coefficients and $C_{21}$ is the collisional
de-excitation rate. This expression differs slightly from the standard
expression \citep[e.g.][]{Madau97}. Here, $Q[J_{\alpha}(\nu)]$ is a function that depends on the exact Ly$\alpha$ spectral shape (see Eq.~\ref{eq:q}), and which classically obtains a value $Q[J_{\alpha}(\nu)]\sim 1$. However, Hirata (2006, his Eq.~B.18) has recently shown that due to a quantum interference term, the probability that a spin-flip occurs through scattering by Ly$\alpha$ wing photons is strongly reduced. Therefore, $Q[J_{\alpha}(\nu)]$ may become very small in cases in which all Ly$\alpha$ photons are in the wings of the line
profile. In this paper we will express our answers as a function of
$Q[J_{\alpha}(\nu)]$, and keep in mind that it may obtain very small
values.

The denominator of Eq.~\ref{eq:tspin} shows that $P_{\alpha} > 27/4(T_{\alpha}/T_*)A_{21}$ $\sim 3 \times 10^{-11}$ s$^{-1}$($T_{\alpha}/[- 100$ K])$(1/Q[J_{\alpha}(\nu)])$ results in $T_s<0$, provided that the Ly$\alpha$ scattering rate dominates the collision rate and the CMB scattering rate. The Ly$\alpha$ scattering rate may be written as 
\begin{equation}
P_{\alpha}=\frac{\sigma_0}{h\nu_0}\frac{L_{{\rm Ly}\alpha}}{4\pi r^2}\int \mathcal{J}(x)\phi(x)dx,
\label{eq:pscat}
\end{equation} where $\mathcal{J}(x)$ is the normalised spectrum ($\int\mathcal{J}(x)dx=1$). We will use this equation to get a constraint on the Ly$\alpha$ luminosity in \S~\ref{sec:subsim}.

\section{The Ly$\alpha$ Color Temperature in a Neutral Collapsing Gas Cloud}
\label{sec:collapse}

We use the Monte-Carlo code described in \citet{Dijkstra06} for our
radiative transfer calculation. This code calculates the Ly$\alpha$
transfer through spherically symmetric gas clouds, with arbitrary density
and velocity fields (for a detailed description the interested reader is referred to Dijkstra et al 2006). For the work presented in this paper the code was expanded to follow the time evolution of the Ly$\alpha$ radiation field as a function of position. This was achieved by keeping track of the total time that elapsed since a Ly$\alpha$ photon was emitted. When this time reaches user-specified values, a 'snapshot' is taken and the locations and frequency distribution of Ly$\alpha$ photons in the frame of the infalling gas are recorded.

\subsection{The Model}
\label{sec:subsim}
We study a Ly$\alpha$ source point source that is embedded in a neutral collapsing gas cloud. The total (dark matter + baryons) mass of the halo hosting the
collapsing cloud is $M_{\rm tot}=10^{11}M_{\odot}$. Collapse occurs at
redshift $z=8$. Within the virial radius, $r_{\rm vir}=16$ kpc the gas density is assumed to trace the dark matter (which is given by a NFW profile with concentration parameter $c=5$)
modified by a core at $r<3 r_s/4$ \citep[see][their Eq. 9-11]{Maller04}
where $r_s=r_{\rm vir}/c$. The infall velocity field, $v(r)$, is assumed to
be linear with ${\rm v}(r)=-v_{\rm circ}[r/r_{vir}]$ in which $v_{\rm
circ}$(=158 km s$^{-1}$) is the circular velocity of the host dark matter
halo. This velocity field corresponds to that of a spherical top-hat model \citep{Dijkstra06}. For realistic initial density profiles, the mean density 
enclosed within radius $r$ decreases smoothly with radius, causing the inner shells to be decelerated more relative to the overall expansion of the
universe. This enhances the infall speed at small $r$ relative to that
in the top-hat model, and results in a flatter velocity profile, possible with an increasing infall speed towards smaller radii. Such a velocity field would result in a larger blue-shift of the Ly$\alpha$ line (see Fig~4 of Dijkstra et al 2006), which would not affect our result regarding the sign of the Ly$\alpha$ color temperature. Beyond the virial radius, $r > r_{\rm vir}$, infall of gas continues and we adopt the density and velocity profiles based on the curves shown in Figures 2 and 4 of Barkana (2004; see Dijkstra et al, 2007, for a more detailed discussion of this choice). 

The gas collapsing outside the galaxy is assumed to be cold, $T_{\rm gas}=300$ K. This low gas temperature reflects the fact that gas in the pre-reionisation universe has a low temperature. Furthermore, calculations by \citet{Birnboim03} have shown that virial shocks are likely not to exist in objects with masses below $M_{\rm tot} \sim 2 \times 10^{11} M_{\odot}$ \citep[also see][]{K05}. In the absence of a virial shock gas accretion occurs 'cold'.  Of course, density inhomogeneities in supersonically collapsing gas will likely cause it to shock heat to a higher temperatures. However, instead of discussing whether our model is realistic or not, we simply point out that the Ly$\alpha$ cross section is independent of gas temperature in the wings of the line profile. Therefore, the scattering process and the Ly$\alpha$ color temperature are practically independent of the assumed gas temperature throughout this paper.

Photons are injected at $t=0$ following a Gaussian frequency distribution
with $\sigma_v=v_{\rm circ}$ \citep[see][for a motivation of this choice]{D07}.
Note that our results do not depend on the precise input spectrum of the Ly$\alpha$ photons, as long as the collapsing cloud is optically thick to all emitted photons.

\subsection{The Results}
\label{sec:result}
\begin{figure}
\vbox{\centerline{\epsfig{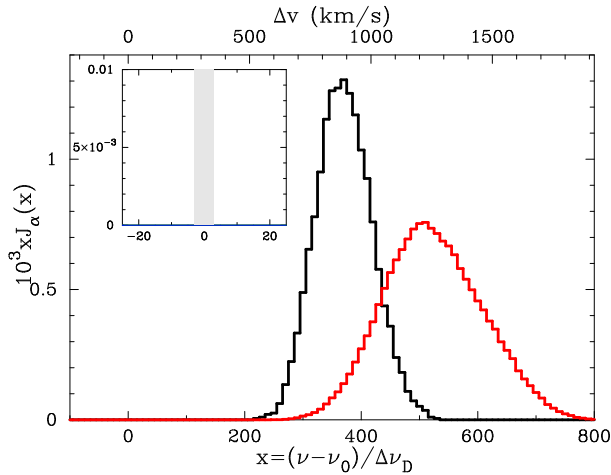}}}
\caption[]{Time evolution of the Ly$\alpha$ spectrum 'seen' by a gas
element at $r=15$ kpc, inside a collapsing cloud of neutral hydrogen. The
Ly$\alpha$ radiative transfer was calculated using a Monte-Carlo code. All
Ly$\alpha$ photons are inserted at $t=0$, and the spectra are shown at
$t=4t_{\rm cross}$ ({\it black}) and $t=10t_{\rm cross}$ ({\it red}) ,
where $t_{\rm cross}\equiv r_{\rm sn}/c$ is the light-crossing time to
radius $r_{\rm sn}=15$ kpc. All spectra are systematically blueshifted
relative to the line center, which is due to energy transfer from the
infalling gas to the Ly$\alpha$ photons. These distributions of Ly$\alpha$
photons have $T_{\alpha}\sim -160$ K ($t=4t_{\rm cross}$) and
$T_{\alpha}\sim -230$ K ($t=10t_{\rm cross}$). When the Ly$\alpha$ scattering rate exceeds, $P_{\alpha} \gsim 5-7 \times 10^{-11}$ s$^{-1}$, the spin temperature may be negative, $T_{s} < 0$.}
\label{fig:sim} 
\end{figure}

 Snapshots were taken of the Ly$\alpha$ frequency distribution at $r_{\rm sn}=5, 10, 15$ and $20$ kpc at four different times: $t=4t_{\rm cross}$, $t=6t_{\rm cross}$, $t=8t_{\rm cross}$ and $t=10t_{\rm cross}$, where $t_{\rm cross}\equiv r_{\rm sn}/c$ is the light-crossing time to radius $r_{\rm sn}$. The spectra 'seen' by the gas at $r=15$ kpc are shown in Figure~\ref{fig:sim} as the {\it black} ($t=4t_{\rm cross}$) and {\it red} ($t=10t_{\rm cross}$) histogram.
The Ly$\alpha$ photons lie systematically to the blue side of the line. The inset of the Figure zooms on the line core and shows more clearly that no
photons are present in the core of the line profile. The reason for the
large shift to the blue side of the line can be easily understood: as the
Ly$\alpha$ photons propagate outward through an optically thick collapsing
gas cloud, energy is transferred from the gas to the photons, which results
in an overall blueshift of the line \citep{Zheng02,Dijkstra06}. This blueshift can exceed the infall velocity of the infalling gas significantly (large velocity shifts are common in Ly$\alpha$ radiative transfer
through optically thick gas clouds. In order for Ly$\alpha$ to escape
from an optically very thick gas cloud, the photons need to diffuse far into
the line wings, where the gas cloud becomes optically thinner).
 
At each subsequent timestep the Ly$\alpha$ line shifts slightly further
to the blue and broadens. This broadening of the line is a direct consequence of the frequency diffusion associated with the Ly$\alpha$ scattering process. Frequency diffusion in a collapsing gas cloud preferentially shifts Ly$\alpha$ photons further to the blue. As we discuss next, the Ly$\alpha$ color temperature is negative at all times.
\subsection{The Ly$\alpha$ Color Temperature and Hydrogen Spin Temperature}
\label{sec:subsim}

Using Eq~(\ref{eq:tcolor}) we find that $T_{\alpha}=-160$ K at $t=4t_{\rm cross}$ and $T_{\alpha}=-230$ K at $t=10t_{\rm cross}$ for a gas shell at $r=15$ kpc (Fig.~\ref{fig:sim}). Therefore, neutral collapsing gas clouds surrounding a central Ly$\alpha$ source may 'see' a negative Ly$\alpha$ color temperature. At later times ($t>10t_{\rm cross}$) the Ly$\alpha$ scattering rate decreases to zero, as all photons have diffused away spatially. For a source that emits Ly$\alpha$ photons at a constant rate, a gas element at $r=15$ kpc would continuously be exposed by a weighted average of the spectra shown in Figure~\ref{fig:sim}, which would clearly also have an associated negative color temperature.
   
For $T_{\alpha}=-200$ K the spin temperature is negative when the Ly$\alpha$ scattering rate exceeds $P_{\alpha} \gsim 5-7 \times 10^{-11}/Q[J_{\alpha}(\nu)]$ s$^{-1}$ (\S~\ref{sec:tspin}). This requires the Ly$\alpha$ luminosity of the central source to exceed a few times $10^{45}/Q[J_{\alpha}(\nu)]$ erg s$^{-1}$ (Eq.~\ref{eq:pscat}). To produce Ly$\alpha$ luminosities of this magnitude requires high star formation rates. If stars form according to the conventional Salpeter mass function, the Ly$\alpha$ luminosity and star formation rate, $\dot{M}_*$, are related by $L_{{\rm Ly}\alpha}\sim 1-3 \times 10^{42}(\dot{M}_*/M_\odot~{\rm yr^{-1}})$ erg s$^{-1}$ \citep[e.g.][]{D07}. For a top-heavy mass function of stars, the Ly$\alpha$ luminosity may be higher by an order of magnitude \citep{Schaerer03}. Therefore, $L_{{\rm Ly}\alpha} \gsim 10^{45}/Q[J_{\alpha}(\nu)]$ erg s$^{-1}$ requires $\dot{M}_* \gsim 10^2/Q[J_{\alpha}(\nu)]$ $M_{\odot}/$yr. Note the importance of quantum interference for this work. As is shown in Appendix~\ref{app:qi}, existing calculations suggest that $Q[J_{\alpha}(\nu)]\sim 10^{-5}$, which implies that unphysically large star formation rates that are a factor of $\sim 10^5$ times larger are required for a population inversion when quantum interference is included. 

A caveat to the previous discussion is that for star formation rates mentioned above, $\dot{M}_* \gsim 10^2/Q[J_{\alpha}(\nu)]$ $M_{\odot}/$yr, the surrounding gas is very unlikely neutral and cold (which was assumed in the radiative transfer calculation). This implies that although the scenario described above may result in a negative Ly$\alpha$ color temperature, the Ly$\alpha$ luminosity that is required to couple the Ly$\alpha$ color temperature to the hydrogen spin temperature is too large. In the next sections, we discuss additional factors that determine whether 21-cm masers can exist.

\section{The Radiation Pressure Exerted by Luminous Ly$\alpha$ Sources}
\label{sec:pressure}

We showed above that Ly$\alpha$ luminosities exceeding $\sim 10^{45}/Q[J_{\alpha}(\nu)]$ erg s$^{-1}$ are required for a population inversion in the hyperfine levels of atomic hydrogen in the collapsing gas. Here we show that the radiation pressure exerted by the Ly$\alpha$ photons is likely large enough to halt the collapse of the cloud. The condition for Ly$\alpha$ radiation pressure to unbind gravitationally collapsed gas is \citep[e.g][]{OH02}
\begin{equation}
L_{{\rm Ly}\alpha}t_{\rm trap}>U_{\rm bind}\sim \frac{GM^2(<r)}{r},
\end{equation} where $U_{\rm bind}$ is the gas' gravitational binding energy, $M(<r)$ is the total mass enclosed in a sphere of radius $r$, and $t_{\rm trap}$ is the total time Ly$\alpha$ photons are 'trapped' inside the sphere of radius r, before escaping. If the gas were optically thin to Ly$\alpha$, then $t_{\rm trap}=r/c$. In optically thick gas, scattering of Ly$\alpha$ photons increases this timescale to $t_{\rm trap}=f(\tau)\times (r/c)$, where $f(\tau)\sim 15(\tau(r)/10^6)^{1/3}(T_{\rm gas}/10^4 \hs {\rm K})^{1/6}$, where $\tau(r)$ is the total Ly$\alpha$ optical depth to radius $r$ \citep{OH02}. For the density profile used in the radiative transfer calculation (\S~\ref{sec:subsim}) we find that radiation pressure can exceed gravity for Ly$\alpha$ luminosities $L_{{\rm Ly}\alpha}\gsim 10^{44}$ erg s$^{-1}$ (at all radii of interest). It is likely that while the central source's Ly$\alpha$ luminosity is building up to the values required to establish a population inversion, the collapse of the cloud has halted. Since gas collapse was essential in the mechanism proposed here to make $T_{\alpha}<0$, we conclude that the required Ly$\alpha$ luminosities are too high (especially for very small values of $Q[J_{\alpha}(\nu)]$, which are favored by the most up-to-date calculations, see \S~\ref{app:qi}).
\section{Constraints on the Properties of the Central Ly$\alpha$ Source} 
\label{sec:sourcereq}
A key requirement for a negative hydrogen spin temperature is the absence of -or more accurately, a low flux of- Ly$\alpha$ photons in the line center, as these would scatter and rapidly rearrange the level populations in the hyperfine transition such that $T_s \rightarrow T_{\rm gas}$. So far, our calculation assumed that all Ly$\alpha$ photons were emitted by the central source. Energy transfer from the gas to the photons caused Ly$\alpha$ photons to blueshift into the line wing on their way out of the collapsing gas cloud, causing a large volume of the cloud to practically see only Ly$\alpha$ wing photons. However, in astrophysically realistic situations Ly$\alpha$ core photons may be injected at any location through different channels. These channels are discussed below. 

\subsection{Alternative Sources of Ly$\alpha$ Core Photons} 
\label{sec:lyacore1}
Different channels that produce Ly$\alpha$ core photons include:

{\bf 1.} Ly$\alpha$ that is produced following photo-excitation by soft UV-photons, defined as photons in the energy range $E=10.2-13.6$ eV. Soft-UV photons can generally penetrate deep into a neutral medium, where they are absorbed by neutral hydrogen atoms. Recombination events produce higher-order Lyman-series photons, which have a $\sim 35\%/5=7\%$ probability of being converted into a Ly$\alpha$ core photon, per scattering event. The factor of 5 reflects the fact that a higher order Lyman-series photon scatters on average $\sim 5$ times before undergoing a cascade (which results in a Ly$\alpha$ photon $\sim 35\%$ of the time, see Pritchard \& Furlanetto 2006).

{\bf 2.} Ly$\alpha$ that is produced following x-ray heating. When hydrogen recombines after being photoionised by x-rays that penetrated the neutral gas, a Ly$\alpha$ photon is produced with a probability of $\sim 2/3$ (for case B recombination). The emergent Ly$\alpha$ photon has a high probability of being injected close to the line center. This effect is probably not important because the recombination time $t_{\rm rec}=1/(n_e\alpha)\gg t_{\rm cross}$, particularly in an almost completely neutral medium (regardless of the gas temperature). A more important effect however, is that photoionisation by x-rays may produce energetic electrons that excite the $n=2$ level of atomic hydrogen, which immediately results in the emission of a Ly$\alpha$ core photon \citep[e.g.][]{Chen06}. 

{\bf 3.} Ly$\alpha$ produced in collisions that excite the Ly$\alpha$ transition. This effect is only important when the gas temperature is close to $10^4$ K.

The newly produced Ly$\alpha$-core photons interact strongly with the neutral gas, and will drive the spin temperature to the gas temperature \citep[as in][]{Deguchi85}. The evolution of the hydrogen spin temperature ultimately depends on which population of Ly$\alpha$ photons dominates the spin-flip rate. If the central source is sufficiently bright in Ly$\alpha$ relative to the soft-UV and x-ray flux, then the rate at which Ly$\alpha$ wing photons produce spin-flips can still dominate. The prominence of the Ly$\alpha$ emitted by a source relative to its continuum is quantified by its equivalent width. Therefore, the requirements mentioned above can be quantified as a constraint on the source's Ly$\alpha$ equivalent width. 

\subsection{Constraints on the Ly$\alpha$ Source's Ly$\alpha$ Equivalent Width}
\label{sec:lyacore2}
The scattering rate of the newly created Ly$\alpha$ core photons at radius $r$ can be estimated as
\begin{equation}
P_{\alpha{\rm, core}}(r)=\langle N \rangle \epsilon(r),
\label{eq:pcon}
\end{equation} Here, $\epsilon(r)$ is the rate at which Ly$\alpha$ core photons are injected at radius $r$. Furthermore, $\langle N \rangle$ is the total number of times a Ly$\alpha$ core photon scatters before it disappears. To estimate $\langle N \rangle$, we first note that it takes on average $\sim 10^4-10^5$ scattering events for a Ly$\alpha$ core photon to be scattered into the wing, $|x|>x_t \sim 3$ \citep{Hummer62}. However, photons close to the core-wing transition frequency, $|x_t|$, can scatter back into the line core in a single scattering event. Therefore, $\langle N \rangle$ should be higher. Velocity gradients facilitate the escape of Ly$\alpha$ photons. In the presence of velocity
gradients the majority of the scattering events of a photon occur around
its resonant region. For a {\it static} medium, a Ly$\alpha$ photon
scatters on average $\sim \tau_0$ times as it traverses an optical depth
$\tau_0$ \citep{Adams72,Harrington73}. The total optical depth of the
resonant region, $\tau_{\rm res}$, can be calculated from the total optical
depth over the scale for which the velocity changes by a thermal width
(also known as the Sobolev length), $v_{\rm th}$: $\tau_{\rm
res}=n_{H}\sigma_0 v_{\rm th}[dv_{\rm infall}/dr]^{-1}$. If we approximate
$dv_{\rm infall}/dr \sim v_{\rm circ}/r_{\rm vir}$, where $v_{\rm circ}$
and $r_{\rm vir}$ are the circular velocity and virial radius of the host
dark matter halo, then $\tau_{\rm res} \sim 2 \times
10^6[(1+\delta)/60][(1+z_{\rm vir})/9]^{3/2}$. Here, $\delta$ is the
overdensity of the gas. Therefore, for a photon to escape from its local
resonant region, it scatters on average $\langle N\rangle \sim \tau_{\rm
res}\sim 10^6-10^7$ times.

In \S~\ref{sec:subsim} we showed that $T_s<0$ required $P_{\alpha} \gsim 5-7 \times 10^{-11}/Q[J_{\alpha}(\nu)]$ s$^{-1}$. Let us require that the scattering rate of Ly$\alpha$ core photons is subdominant, e.g. $P_{\alpha{\rm, core}}(r) < 10^{-11}$ s$^{-1}$ (the factor $Q[J_{\alpha}(\nu)]$ is deliberately omitted, as quantum interference in the core can be ignored). Then, Eq~\ref{eq:pcon} implies that $\epsilon(r)<10^{-11}$ s$^{-1}$/$\langle N \rangle$. 

The rate at which soft-UV photons are converted into Ly$\alpha$ is given by
\begin{equation}
\epsilon(r)\sim 0.07\times 4\pi \int_{{\rm Ly}\gamma}^{{\rm Lyc}} \frac{J(\nu)}{h\nu}\sigma_{\rm tot}(\nu) e^{-\tau(\nu,r)} d\nu.
\label{eq:epsr}
\end{equation} 
Here, the factor $0.07$ is the mean probability for a higher-order Lyman-series photon to be converted per scattering \citep{PF06a}, $\tau(\nu,r)=N_{H}(r)\sigma_{\rm tot}(\nu)$, in which $\sigma_{\rm tot}(\nu)$ is the absorption cross section at frequency $\nu$ which is obtained by summing over all Lyman series transitions (derived in Appendix~\ref{app:lyn}), and $N_H(r)$ is the total column of neutral hydrogen between the source and radius $r$. The integral is performed over $h\nu \in [12.8-13.6]$ eV, as Ly$\beta$ photons cannot be converted into Ly$\alpha$ photons \citep{PF06a}. We assume that the flux density is constant, i.e. $J(\nu)=K$, where $K$ is a constant, and that $N_H(r)=10^{21}$ cm$^{-2}$ and numerically evaluate the integral. The constraint $\epsilon(r)<10^{-11}$ s$^{-1}$/$\langle N \rangle$ then translates to $K<1.5\times 10^{-14} $/$\langle N \rangle$ ergs s$^{-1}$ cm$^{-2}$ sr$^{-1}$ Hz$^{-1}$, which can be recast as $16\pi^2 r^2 K< 10^{46}$/$\langle N \rangle$ ergs s$^{-1}$ \AA$^{-1}$. Since the Ly$\alpha$ luminosity of the central source was required to be $L_{{\rm Ly}\alpha}\gsim 10^{45}/Q[J_{\alpha}(\nu)]$ ergs s$^{-1}$, this requires that the Ly$\alpha$ equivalent width is EW$\gsim 0.1\langle N \rangle/Q[J_{\alpha}(\nu)]$ \AA$= 10^6/Q[J_{\alpha}(\nu)]$ \AA, for $\langle N \rangle=10^7$. Although the precise lower limit depends on model parameters such as $N_{\rm HI}$ and $\langle N \rangle$, our lower limit on EW provides a reasonable order-of-magnitude estimate.

For comparison, a normal population can theoretically produce a maximum of EW=200-300\AA\hs \citep[e.g.][]{Schaerer03}. This number can be higher by about a factor of $5$ for population III stars and quasars, but this is still well below what is required to produce a 21-cm maser. However, it has been shown that a multiphase interstellar medium (ISM) can preferentially transmit Ly$\alpha$ photons over continuum photons \citep{Neufeld91,Hansen06}. In multiphase ISM models, the ISM consists of cold, dense, dusty clumps embedded in a hot medium. The Ly$\alpha$ scatters off the surface of the cold clumps and propagates mainly through the hot, dust free, interclump medium. On the other hand, the continuum photons penetrate the clumps where it is subject to absorption by dust. Therefore, a multiphase ISM can preferentially absorb the continuum flux and boost the Ly$\alpha$ EW. It is currently not known whether this effect actually occurs in nature, especially at the level that is required to make EW$\gsim 10^6/Q[J_{\alpha}(\nu)]$ \AA. We point out that the multi-phase ISM that is required to produce EW$\gsim 10^6/Q[J_{\alpha}(\nu)]$ \AA\hs is likely also sufficient to counteract the negative impact of x-rays. For a spectrum of the form $f_{\nu}\propto \nu^{-\alpha}$ the injection rate of Ly$\alpha$ photons through x-rays is comparable to that given by Eq~(\ref{eq:epsr}) for $\alpha >2$. 

\section{Conclusions}
\label{sec:conclusion}
We have performed Monte-Carlo calculations of the radiative transfer of Ly$\alpha$ photons emitted by a source embedded in a neutral collapsing gas cloud and computed the Ly$\alpha$ spectrum as function of radius and time. This source represents a young galaxy or quasar during the early stages of the epoch of reionisation (EoR). After deriving an expression for the Ly$\alpha$ color temperature $T_{\alpha}$ for arbitrary frequency distributions of Ly$\alpha$ photons (Eq.~\ref{eq:tcolor}), we show that neutral collapsing gas clouds surrounding a central Ly$\alpha$ source may see a negative Ly$\alpha$ color temperature (Fig~\ref{fig:sim}) under certain circumstances.

We have investigated whether a population inversion in the hyperfine transition of atomic hydrogen via the Wouthuysen-Field effect can be established in such scenarios, thus producing a 21-cm maser. The reason for this investigation is clear: if 21-cm masers exist during the epoch of reionisation (EoR), this could greatly boost the expected 21-cm emissivity of neutral gas during the EoR. Unfortunately, we have found that in practice a population inversion does not occur for several reasons:

(1) The required Ly$\alpha$ luminosities exceed $\sim 10^{45}$ erg s$^{-1}$. Although this can be obtained with physically possible star formation rates ($\dot{M}_*\gsim 10^2M_{\odot}$ yr$^{-1}$) the radiation pressure exerted by such a large Ly$\alpha$ flux likely halts the collapse of the cloud (\S~\ref{sec:pressure}).

(2) Quantum interference can strongly reduce the strength of the WF-effect in the Ly$\alpha$ line wings \citep{Hirata06}. The probability that a spin-flip occurs when a Ly$\alpha$ wing photon scatters is reduced by a factor of $Q=(x_{45}/x)^2$ (see Appendix~\ref{app:qi}) when quantum interference is included. In the context of this paper, this factor can be as small as $Q=10^{-4}-10^{-6}$. Therefore, when this quantum correction to the WF-coupling strength is applied, the required Ly$\alpha$ luminosities are larger by orders of magnitude. 

(3) A key requirement for having $T_{\alpha}<0$ is a very low level of Ly$\alpha$ photons in the line center, as these would scatter and rapidly rearrange
the level populations in the hyperfine transition such that $T_s \rightarrow T_{\rm gas}$. We discussed in \S~\ref{sec:sourcereq}, however, that Ly$\alpha$ core photons can be injected at a given location via different channels including x-ray heating, collisional excitation and conversion of higher Lyman series photons. The negative impact of these newly produced Ly$\alpha$ photons (both by x-rays and soft-UV photons) is only overcome when the central source has an extremely prominent Ly$\alpha$ line with EW$\gsim 10^6/Q$ \AA, which is several orders of magnitude larger than that of currently known sources in the Universe.

Unless the high redshift Universe contains some truly remarkable sources that can overcome the three obstacles mentioned above simultaneously, the Wouthuysen-Field effect will not produce a population inversion in the 21-cm transition. Detecting the redshifted 21-cm from the epoch of reionisation will therefore remain among the greatest observational challenges in cosmology.

{\bf Acknowledgments} We thank George Rybicki and Stuart Wyithe for useful discussions. MD would like to thank Harvard's Institute for Theory \& Computation for its hospitality. We thank an anonymous referee for providing us with some useful references to earlier work on 21-cm Masers. MD is supported by the Australian Research Council. AL was supported in part by NASA grants NAG 5-1329 and NNG05GH54G.

\onecolumn
\appendix
\section{Derivation of the Ly$\alpha$ Color and Hydrogen Spin Temperature }
\subsection{The 6-Level Hydrogen Atom without Collisions and CMB photons}
\label{app:6level}
In this section we derive the expression for the Ly$\alpha$ color
temperature taking into account hyperfine splitting in the Hydrogen atom's $n=1$ and $n=2$ states, following Meiksin (2000).
\begin{figure}
\vbox{ \centerline{\epsfig{file=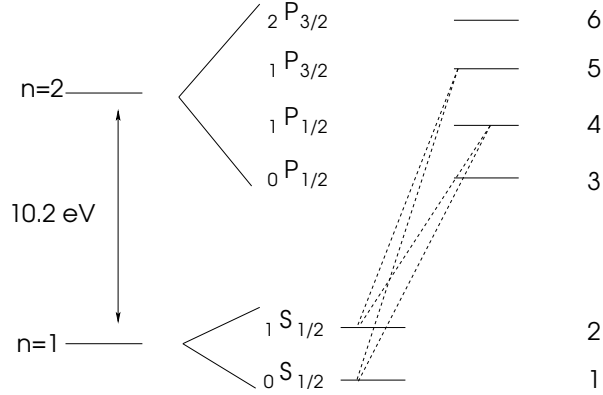,width=8.0truecm}}}
\caption[6levelatom]{Schematic depiction of hyperfine splitting of the electronic ground state (n=1) and the first excited state (n=2) of the hydrogen atom. This 6-level model has been used to arrive at the expressions for the Ly$\alpha$ color temperature, Eq.~\ref{eq:talpha2} and the spin temperature, Eq.~\ref{eq:tspin2}. The {\it dashed lines} denote the transitions involved in the mixing
of hyperfine levels. Levels 3 and 6 do not contribute to the mixing of levels. The notation used for the individual levels is $_{F}L_{J}$, where $L$ is the orbital angular momentum, $J$ the total angular momentum quantum number. Here, $F$ is the sum of $J$ and $I$, the nuclear spin quantum number.}
\label{fig:6level} 
\end{figure}
Ignoring collisions and absorption and re-emission of CMB photons, the rate equation for atoms in the $n=1$ state is
\begin{eqnarray}
\frac{\partial n_1}{\partial t}=n_2 A_{21} +\bar{J}\Big{[} 
-B_{14}\frac{A_{42}}{A_{42}+A_{41}}n_1+B_{24}\frac{A_{41}}{A_{41}+A_{42}}n_2-B_{15}\frac{A_{52}}{A_{52}+A_{51}}n_1+B_{25}\frac{A_{51}}{A_{51}+A_{52}}n_2\Big{]}=\nonumber \\
=n_2 A_{21} +\bar{J}\frac{c^2}{2h\nu_0^3}\Big{[} 
-\frac{g_4}{g_1}\frac{A_{41}A_{42}}{A_{42}+A_{41}}n_1+\frac{g_4}{g_2}\frac{A_{42}A_{41}}{A_{41}+A_{42}}n_2-\frac{g_5}{g_1}\frac{A_{51}A_{52}}{A_{52}+A_{51}}n_1+\frac{g_5}{g_2}\frac{A_{52}A_{51}}{A_{51}+A_{52}}n_2\Big{]}
\end{eqnarray} 
where we used the relations $B_{du}=\frac{g_u}{g_d}B_{ud}=\frac{g_u}{g_d}
\frac{c^2}{2h\nu_0^3}A_{ud}$, and the notation $\bar{J}\equiv \int
J(\nu)\phi(\nu) d\nu$. Following \citet{Meiksin00},
\begin{equation}
A_{41}=\frac{1}{3}A_{\alpha}, \hs \hs 
A_{42}=\frac{2}{3}A_{\alpha}, \hs \hs
A_{51}=\frac{2}{3}A_{\alpha}, \hs \hs
A_{52}=\frac{1}{3}A_{\alpha}, \hs \hs, 
\end{equation} where $A_{\alpha}=6.25 \times 10^8$ s$^{-1}$, is the total Einstein A-coefficient, and $g_1=1$, $g_2=3$, $g_4=3$ and $g_5=3$. We get
\begin{eqnarray}
\frac{\partial n_1}{\partial t}=n_2 A_{21}+\frac{c^2}{2h\nu_0^3}A_{\alpha}\times \\ \nonumber
\Big{[} -\frac{2}{3}n_1\int J(\nu)\phi_{14}(\nu)d\nu+\frac{2}{9}n_2\int J(\nu) \phi_{24}(\nu)d\nu-\frac{2}{3}n_1\int J(\nu)\phi_{15}(\nu)d\nu+\frac{2}{9}n_2\int J(\nu) \phi_{25}(\nu)d\nu\Big{]}.
\label{eq:rate}
\end{eqnarray}  All line profiles are identical, apart from an offset in frequency: $\phi_{24}(\nu)=\phi_{14}(\nu+\Delta \nu_S)$,  $\phi_{25}(\nu)=\phi_{15}(\nu+\Delta \nu_S)$ and  $\phi_{14}(\nu)=\phi_{15}(\nu+\Delta \nu_G)$, where $\Delta \nu_G=10.89$ Ghz  and $\Delta \nu_S=1.42$ Ghz \citep{Hirata06}. If we approximate $\phi(\nu+\Delta \nu)\approx \phi(\nu)+\Delta \nu\frac{\partial \phi}{\partial \nu}$ and write everything in terms of $\phi_{15}$ (which is written as $\phi(\nu)$ hereafter), and replace $n_2$ with $n_H-n_1$, we get
\begin{eqnarray}
\frac{\partial n_1}{\partial t}=n_2A_{21}+\frac{c^2}{9h\nu_0^3}A_{\alpha}\times \\ \nonumber
\Big{[}n_H\int J(\nu)\Big{(}2\phi(\nu)+\frac{\partial \phi}{\partial \nu}\Delta \nu_G+2\frac{\partial \phi}{\partial \nu}\Delta \nu_S
\Big{)}d\nu-n_1\int J(\nu)\Big{(}8\phi(\nu)+4\frac{\partial \phi}{\partial \nu}\Delta \nu_G+2\frac{\partial \phi}{\partial \nu}\Delta \nu_S\Big{)}d\nu
\Big{]}.
\label{eq:bonjovi}
\end{eqnarray} This can be written in the following more compact form
\begin{equation}
\frac{\partial y}{\partial t}=(1-y) A_{21}+b_1-b_2-y(4b_1-b_2)
\label{eq:b}
\end{equation} after dividing the left and right hand side of Eq~\ref{eq:bonjovi} by $n_H$ and after writing $y=n_1/n_H$ and where we defined
\begin{equation}
b_1 \equiv A_{\alpha} \frac{c^2}{9h\nu_{0}^3} \int J(\nu) \Big{(}2\phi(\nu)+\frac{\partial \phi}{\partial \nu}\Delta \nu_G\Big{)}d\nu; \hs \hs \hs
b_2 \equiv -A_{\alpha} \frac{2c^2}{9h\nu_{0}^3} \Delta \nu_S \int \frac{\partial \phi}{\partial \nu}J(\nu)d\nu
\label{eq:b1}
\end{equation} The equilibrium value of $y$ is obtained by setting $\partial y/\partial t$=0 and corresponds to a spin temperature $T_s$ of
\begin{equation}
T_s\approx \frac{3y}{4y-1}T_*=-\frac{b_1}{b_2}T_{*}=\Big{[}\frac{\int \phi(\nu) J(\nu)d\nu}{\Delta \nu_S\int \frac{\partial \phi}{\partial \nu}J(\nu)d\nu}+\frac{\Delta \nu_G}{\Delta \nu_S}\Big{]}T_*\approx \Big{[}\frac{\int \phi(\nu) J(\nu)d\nu}{\Delta \nu_S\int \frac{\partial \phi}{\partial \nu}J(\nu)d\nu}\Big{]}\equiv T_{\alpha},
\label{eq:talpha2}
\end{equation} 
 where we used that $T_{*} \ll T_{\rm s}$ \footnote{At first glance, it is not obvious that this inequality is satisfied. The spin temperature is usually $T_{\rm s}\gg T_*$, while in this paper $T_{\rm s} \ll T_*$. One may therefore expect that there is a regime in which $T_{\rm s}\sim 0$. However, this is not the case. The main reason is that continuously increasing the number of atoms in the $n=2$ state, say from a state in which $T_{\rm s}=T_{\rm gas}$, increases $T_{\rm s}$, where in the limit $n_2 \rightarrow 3n_{\rm H}/4$ one gets $T_{\rm s}\rightarrow \infty$. Increasing the fraction of atoms in the $n=2$ state further results in a change in the sign of the spin temperature and $T_{\rm s}\rightarrow -\infty$. By further increasing the number of atoms in the $n=2$ state, one reduces the absolute value of $T_s$. In practice, the limit $T_{\rm  s}\sim 0$ is never reached as it would correspond to a case in which all atoms are in the $n=2$ state, i.e. $n_{2}\rightarrow n_{\rm H}$.}. Furthermore, we have used that the second term within the square brackets is negligible compared to the first term for all applications in this paper. Here the last definition only states that if Ly$\alpha$ scattering were the dominant process regulating the level populations in the 21 cm transition, then the equilibrium spin temperature is equal to the Ly$\alpha$ color temperature. Since $h\Delta \nu_S=k_BT_*$, this yields Eq.~(\ref{eq:tcolor}).

\subsection{The 6-Level Hydrogen Atom with Collisions and CMB photons}
\label{app:6levelcoll}

In the presence of CMB photons and when collision are accounted for,
Eq.~(\ref{eq:b}) changes to
\begin{eqnarray}
\frac{\partial y}{\partial t}=(1-y)\Big{(}A_{21}+C_{21}+B_{21}I_{\rm CMB}\Big{)}+(b_1-b_2)-y\Big{(}4b_1-b_2+C_{12}+B_{12}I_{\rm CMB}\Big{)}.
\end{eqnarray} In this case the equilibrium solution (obtained by setting $\partial y/\partial t$=0) corresponds to a spin temperature $T_s$ of
\begin{equation}
T_s\approx\frac{3y}{4y-1}T_*=\frac{T_*+\frac{C_{21}}{A_{21}}T_*+\frac{B_{21}}{A_{21}}I_{\rm CMB}T_*+\frac{b_1-b_2}{A_{21}}T_*}{1-\frac{b_2}{A_{21}}+\frac{C_{21}T_*}{A_{21}T_{\rm gas}}}=\frac{T_*+\frac{C_{21}}{A_{21}}T_*+\frac{B_{21}}{A_{21}}I_{\rm CMB}T_*+\frac{b_1(1+\frac{T_*}{T_{\alpha}})T_*}{A_{21}}}{1+\frac{b_1 T_*}{A_{21}T_{\alpha}}+\frac{C_{21}T_*}{A_{21}T_{\rm gas}}}, 
\label{eq:tspin2} 
\end{equation} where we used that $B_{12}=3B_{21}$ and $C_{12}=3C_{21}e^{-T_*/T_{\rm gas}}$ $\approx 3C_{21}(1-T_*/T_{\rm gas})$.  When the Ly$\alpha$ scattering rate is expressed as
\begin{equation}
P_{\alpha}=4\pi \int \frac{J(\nu)}{h\nu} \sigma(\nu) d\nu \approx
\frac{4\pi}{h\nu_0} \frac{B_{\alpha} h\nu_0}{4\pi}\int J(\nu) \phi(\nu) d\nu=
3A_{\alpha}\frac{c^2}{2 h\nu_0^3}\int J(\nu) \phi(\nu) d\nu\sim\frac{27}{4}b_1,
\end{equation} the standard equation for the hydrogen spin temperature \citep[e.g.][]{Madau97} is recovered:
\begin{equation}
T_s=\frac{T_*+T_{\rm cmb}+y_\alpha T_{\alpha}+y_cT_{\rm gas}}{1+y_{\alpha}+y_c},\hspace{5mm}, y_{\alpha}=\frac{4P_{\alpha}T_{*}}{27A_{21}T_{\alpha}}, \hspace{5mm}, y_{c}=\frac{C_{21} T_*}{A_{21} T_{\rm gas}},
\end{equation} where $T_*$ in the numerator is usually omitted as it is negligibly small compared to the other terms in most applications.

\subsection{Expressions of $T_{\alpha}$ and $T_s$ with Quantum Interference}
\label{app:qi}
\citet{Hirata06} has performed detailed quantum mechanical calculations of the cross-section of the process in which absorption and re-emission of Ly$\alpha$ photons results in a transition between levels 1 and 2 (Fig~\ref{fig:6level}). These cross-sections are denoted as $\sigma_{12}$ and $\sigma_{21}$ and represent the transitions $1\rightarrow 2$ and $2\rightarrow 1$, respectively. They are given by
\begin{equation}
\sigma_{12}=\frac{2}{9}\sigma_0[\phi_{14}+\phi_{15}-2\phi_{15/14}],\hspace{5mm}\sigma_{21}=\frac{2}{27}\sigma_0[\phi_{24}+\phi_{25}-2\phi_{25/24}].
\label{eq:sig12}
\end{equation} Here, $\sigma_0$ is the Ly$\alpha$ cross-section evaluated at resonance and $\phi_{15/14}$ and $\phi_{25/24}$ are quantum interference terms\footnote{Note that when the quantum interference terms are ignored the rate equation for $n_1$ can be written as $\frac{\partial n_1}{\partial t}=n_2(A_{21}+P_{21})-n_1P_{12}$, where $P_{12/21}$ is the rate at which Ly$\alpha$ scattering puts atoms from the $1\rightarrow 2/2\rightarrow 1$ state. This can be written as$\frac{\partial n_1}{\partial t}=n_2(A_{21}+4\pi n_2\int \frac{J(\nu)}{h\nu}\sigma_{21}(\nu)d\nu-4\pi n_1\int \frac{J(\nu)}{h\nu}\sigma_{21}(\nu)d\nu$. If we use Eq~\ref{eq:sig12} and ignore the quantum interference term, we are left with Eq~\ref{eq:rate}. Therefore, apart from the quantum interference term, our calculations are in exact agreement with those presented by \citet{Hirata06}.} which are defined as
\begin{equation}
\phi_{15/14}(\nu)=\frac{1}{\sqrt{2\pi}\sigma_{\nu}}\int d\nu' e^{-(\nu-\nu')^2/2\sigma_\nu^2}\phi^u_{15/14}(\nu');\hspace{5mm}\phi^u_{15/14}(\nu)=\frac{\gamma}{\pi}\frac{\Delta \nu_{14}\Delta \nu_{15}+\gamma^2}{\Big{[}\Delta \nu_{14}^2+\gamma^2\Big{]}\Big{[}\Delta \nu_{15}^2+\gamma^2\Big{]}},
\label{eq:qi}
\end{equation} (and $\phi_{25/24}$ is defined equivalently) where $\sigma_{\nu}=\sqrt{\frac{k_B T_{\rm gas}}{m_p c^2}}\nu_0$, in which $m_p=1.66\times 10^{-24}$ g is the proton mass. Furthermore, $\gamma\equiv A_{\alpha}/4\pi$, $\Delta \nu_{15}=(\nu-\nu_{15})$, where $\nu_{15}$ is the frequency at resonance of the $1\rightarrow 5$ transition \citep[][the other symbols are defined similarly]{Hirata06}. For comparison, a normal 'unconvolved line profile' is given by a Lorentzian $\phi^u_{15}(\nu)=\frac{\gamma}{\pi}\frac{1}{\Delta \nu_{15}^2+\gamma^2}$ \citep[e.g.][Eq~10.73]{RL79}. For Ly$\alpha$ scattering in the line core, the quantum interference term is typically smaller than the traditional terms and it can be ignored without introducing a large error. However, this is not true in the far line wings. Here, by combining Eq~\ref{eq:sig12} and Eq~\ref{eq:qi} the line profiles $\phi_{12}(x)$ and $\phi_{21}(x)$ can be written as 
\begin{equation}
 \phi_{12}=\frac{4}{9}\frac{a}{\sqrt{\pi}x^2}\Big{(}\frac{x_{45}}{x}\Big{)}^2, \hspace{5mm} \phi_{12}=\frac{4}{27}\frac{a}{\sqrt{\pi}x^2}\Big{(}\frac{x_{45}}{x}\Big{)}^2, \hspace{5mm} \mbox{with quantum interference},
\end{equation} where $x_{45}=(\nu_{14}-\nu_{15})/\Delta \nu_D$. For comparison
\begin{equation}
\phi_{12}=\frac{4}{9}\frac{a}{\sqrt{\pi}x^2}, \hspace{5mm}  \phi_{21}=\frac{4}{27}\frac{a}{\sqrt{\pi}x^2}, \hspace{5mm}\mbox{without quantum interference}.\nonumber
\end{equation} Therefore, the probability that a spin-flip occurs when a Ly$\alpha$ wing photon scatters is reduced by a factor of $(x_{45}/x)^2$ when quantum interference is included. In this paper, this factor can be as small as $10^{-4}-10^{-6}$! We caution that the exact degree at which quantum interference reduces the spin-flip probability is uncertain. The main reason for this uncertainty is that the unconvolved line profiles may deviate from Lorentzian in the far line wings, which may affect these calculations (Zygelman, private communication). 
 
Regardless of whether quantum interference is included, we can write $\phi_{21}(\nu)\equiv \sigma_{21}(\nu)/\sigma_0=\frac{1}{3}\phi_{12}(\nu+\Delta \nu_S)$. By repeating the analysis of \S\ref{app:6level}, one can show that the expression for the Ly$\alpha$ color remains the same. However, in the expression for the hydrogen spin temperature $y_{\alpha}\rightarrow y_{\alpha}Q[J_{\alpha}(\nu)]$ where
\begin{equation}
Q[J_{\alpha}(\nu)]=\frac{\int J_{\alpha}(\nu)[\phi_{24}+\phi_{25}-2\phi_{25/24}]d\nu}{\int J_{\alpha}(\nu)[\phi_{24}+\phi_{25}]d\nu}.
\label{eq:q}
\end{equation} This equation simply states that the cross-section that includes quantum interference should replace the cross-section that ignores this term in the calculation of the spin-flip rate. Note that if all Ly$\alpha$ were far in the line wings, then $Q[J_{\alpha}]\propto x^{-2}$.
 
\section{The Opacity of Gas to Photons in the Energy Range $E=10.2-13.6$ eV.}
\label{app:lyn}
The cross section for Ly$\alpha$ absorption can be written as
\begin{equation}
\sigma_{\alpha}(x)=\sigma_{\alpha,0}\phi(x;A_{21})=f_{12}\frac{\pi e^2}{mc}\frac{1}{\sqrt{\pi}\Delta \nu_0}\phi(x;A_{21}),
\end{equation} where $\sigma_{\alpha,0}$ is the line center cross section, $f_{12}=0.4167$ the Ly$\alpha$ oscillator strength and $\phi(x;A_{21})$ is the Voigt function \citep[][p. 288, Eq. 10.70]{RL79}. The reason for writing the voigt function as $\phi(x;A_{21})$ is that it explicitly  demonstrates that it depends on the Einstein A-coefficient (in the wing of the line profile, $\phi(x)=a/[\sqrt{\pi}x^2]$, where $a=A_{21}/4 \pi \Delta \nu_D$). Similarly, the cross section for higher order Lyman-series photons can be expressed as:
\begin{equation}
\sigma_{n}(x_n)=f_{1n}\frac{\pi e^2}{mc}\frac{1}{\sqrt{\pi}\Delta \nu_{n}}\phi(x_n;A_{n1}).
\end{equation} 
Here, $x_n=(\nu-\nu_n)/\Delta \nu_n$. To compute the cross sections
therefore requires knowledge of the oscillator strengths, $f_{1n}$, and
the Einstein coefficients, $A_{n1}$. \citet{Menzel35} derived the
oscillators strengths \citep[Eq. 10.46 in][]{RL79} to be
\begin{equation}
f_{1n}=\frac{2^9n^5(n-1)^{2n-4}}{3(n+1)^{2n+4}}.
\label{eq:f1n}
\end{equation} From the definition of the oscillator strength we know that
\begin{equation}
\frac{\pi e^2}{mc}f_{1n}=\frac{h\nu_n}{4\pi}B_{1n}=\frac{g_n}{g_1}\frac{h\nu_n}{4\pi}B_{n1}=\frac{g_n}{g_1}\frac{h\nu_n}{4\pi}A_{n1}\frac{c^2}{2h\nu_n^3}
\label{eq:a1n}
\end{equation} 
\citep[][Eq. 10.70]{RL79}. 
A combination of Eqs.~(\ref{eq:f1n}) and (\ref{eq:a1n}) yields $A_{1n}$. 

\begin{figure}
\vbox{\centerline{\epsfig{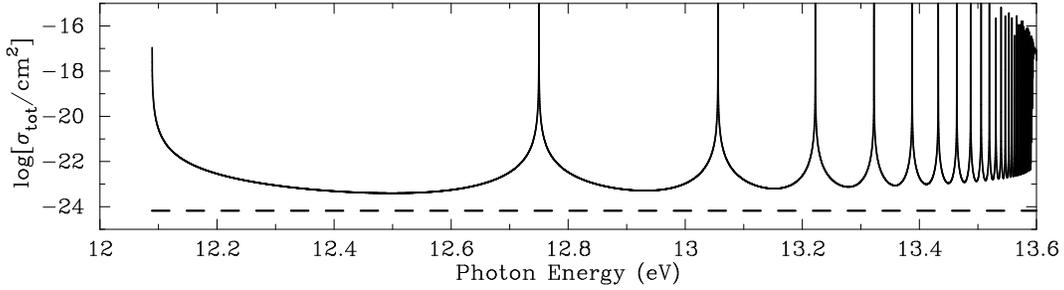}}}
\caption[]{The absorption cross section of neutral atomic hydrogen gas to
soft UV photons. The spikes at E=$12.1$, 12.8, 13.05, ... correspond to the
Ly$\beta$, Ly$\gamma$, Ly$\delta$,...-resonances, respectively. The
horizontal dashed line corresponds to the Thomson cross-section. We assumed that $T_{\rm gas}=300$ K, but note that the gas temperature barely affects this plot (see text).}
\label{fig:lw} 
\end{figure}

Given the cross-sections for the various Lyman-series transitions, it is
possible to calculate the opacity of neutral gas for a photon of any given
energy in the range $E=10.2-13.6$ eV, by summing over all Lyman-series
transitions
\begin{equation}
\sigma_{\nu}=\sum_{i=1}^{\infty}\sigma_i(x_i), \hs\hs x_i=(\nu-\nu_i)/\Delta \nu_i.
\end{equation} 
The result of this calculation is shown in Figure~\ref{fig:lw}. Here, we assumed that $T_{\rm gas}=300$ K. However, the appearance of this plot is very insensitive to the gas temperature, as this only affects the heights and widths of the resonance spikes in very narrow frequency intervals.

\label{lastpage}
\end{document}